\newcommand{\AmS}{{\protect\the\textfont2
  A\kern-.1667em\lower.5ex\hbox{M}\kern-.125emS}}
\newcommand{\beq}{\begin{equation}}
\newcommand{\beqa}{\begin{eqnarray}}
\newcommand{\bce}{\begin{center}}
\newcommand{\bfig}{\begin{figure}}
\newcommand{\bit}{\begin{itemize}}
\newcommand{\ben}{\begin{enumerate}}
\newcommand{\eeq}{\end{equation}}
\newcommand{\eeqa}{\end{eqnarray}}
\newcommand{\ece}{\end{center}}
\newcommand{\efig}{\end{figure}}
\newcommand{\eit}{\end{itemize}}
\newcommand{\een}{\end{enumerate}}
\newcommand{\npb}{Nucl. Phys. {\bf B}}

\documentclass[epj]{svjour}
\usepackage{graphics}
\begin{document}
\title{Theoretical Aspects of Higgs Physics at the 
LHC}
\author{Mauro Moretti \inst{1} 
\thanks{I acknowledge the financial support of the European Union
under the contract HPRN-CT-2000-00149}%
}                     
\institute{Department of Physiscs, Ferrara University and INFN Ferrara}
\date{Received: date / Revised version: date}
%
\abstract{
The strategies recently developed to study Higgs boson properties 
at the LHC are reviewed. It is shown how to obtain model-independent 
determinations of couplings to fermions and gauge bosons by 
exploiting different production and decay channels. 
We consider the case of  Weak Boson Fusion
Higgs production with $H \to b \bar b$ %
} 
\maketitle
\section{Introduction}
The LHC will allow  the discovery of the Higgs boson and the 
study of its mass, width and couplings to 
fermions and gauge bosons. While the decay channels $H \to \gamma \gamma$ 
and $H \to Z Z^{(*)} \to 4l$ will allow a direct mass measurement at 
the 0.1\% level over a wide range of masses~\cite{atl99-15}, 
the total width can only 
be determined with about 10\% accuracy by direct measurement of the decay 
$H \to Z Z^{(*)} \to 4l$ for $m_H > 200$~GeV, (the Higgs width 
for lower Higgs masses being too small with respect 
to the detector resolution). 
An indirect measurement of the total 
width can be performed also in the low mass region  
by exploiting the available production and decay 
mechanisms at the LHC. 
Several studies have been performed to improve on the strategy 
originally proposed in ref.~\cite{zknrw00} for the determination of the 
Higgs boson properties.
We will briefly review the progress recently made in this field.
The main focus will be on the 
mass window 115-200~GeV, the one preferred by electroweak 
precision data and by supersymmetry. 
\section{Theoretical calculations}
In order to disentangle a signal from backgrounds, 
a good understanding of uncertainties in theoretical predictions is necessary. The QCD corrections, at least at next-to-leading order (NLO), are 
known for all production channels, the most recent calculations being the 
NNLO calculation\footnote{The K factor for the NLO contribution
is 2 making desirable the knowledge of NNLO contribution} for the gluon fusion process in the limit of heavy top-quark
mass~\cite{grazzini,kilgore} and the NLO corrections for the process 
$pp / p \bar p \to t \bar t H + X$~\cite{tthnlo}.
At present, the uncertainties arising from QCD uncertainties (combining the 
residual scale dependence with the error from parton distribution functions) 
can be estimated to be  $\pm$20\% for gluon 
fusion, $\pm$5\% for 
WBF, $\pm$10\% for associated production. 

Concerning the backgrounds, several processes with low final state parton
multiplicity (corresponding to important irreducible backgrounds) are available 
at NLO, namely 
$q \bar q \to \gamma \gamma$~\cite{diphox}, 
$g g \to \gamma \gamma$~\cite{gluglugg}, 
$p p(\bar p) \to W b \bar b$, $p p(\bar p) \to Z b \bar b$~\cite{evc}, 
$p p(\bar p) \to W j j$, $p p(\bar p) \to Z j j$~\cite{ce}, 
$p p(\bar p) \to VV$~\cite{fw} and 
QCD $H + jj$ production via gluon fusion~\cite{qcdhjj}. Some of these 
calculations are already implemented in NLO Monte Carlo programs. 
In the case of multiparton final states, the methods developed  
for NLO calculations cannot be applied, because of the complexity of the 
calculations for processes with many external legs. Recently some 
effort 
has been devoted to the realization of LO Monte Carlo event generators 
based on exact matrix element calculations,  interfaced to the shower 
evolution Monte Carlo programs producing the real final state 
hadrons~\cite{alpgen,comphep,grace,madgraph,acermc}.

\section{Higgs couplings to fermions and gauge bosons}
\label{sec:hcoup}
In principle, the Higgs coupling to a given fermion family $f$, 
could be obtained from the following relation:
\begin{eqnarray}
R(H\to f \bar f)&=&\int{L dt}\cdot \sigma(pp\to H)\cdot 
{\frac{\Gamma_f}{\Gamma}}, \nonumber
\end{eqnarray}
where $R(H\to f \bar f)$ is the Higgs production rate in a given final state, 
$\int{L dt}$ is the integrated 
luminosity, $\sigma(pp\to H)$ is the Higgs production cross section, 
while $\Gamma$ and $\Gamma_f$ are the total and partial Higgs widths 
respectively.
Aiming at model-independent coupling determinations, one needs to 
consider ratios of couplings, which are experimentally accessible
through the measurements of ratios of rates for different final
states, because the total Higgs cross-section and width
cancel in the ratios 
(as well as the luminosity and all the 
QCD uncertainties related to the initial 
state). 


\bfig[htb]
\resizebox{0.35\textwidth}{!}{%
  \includegraphics{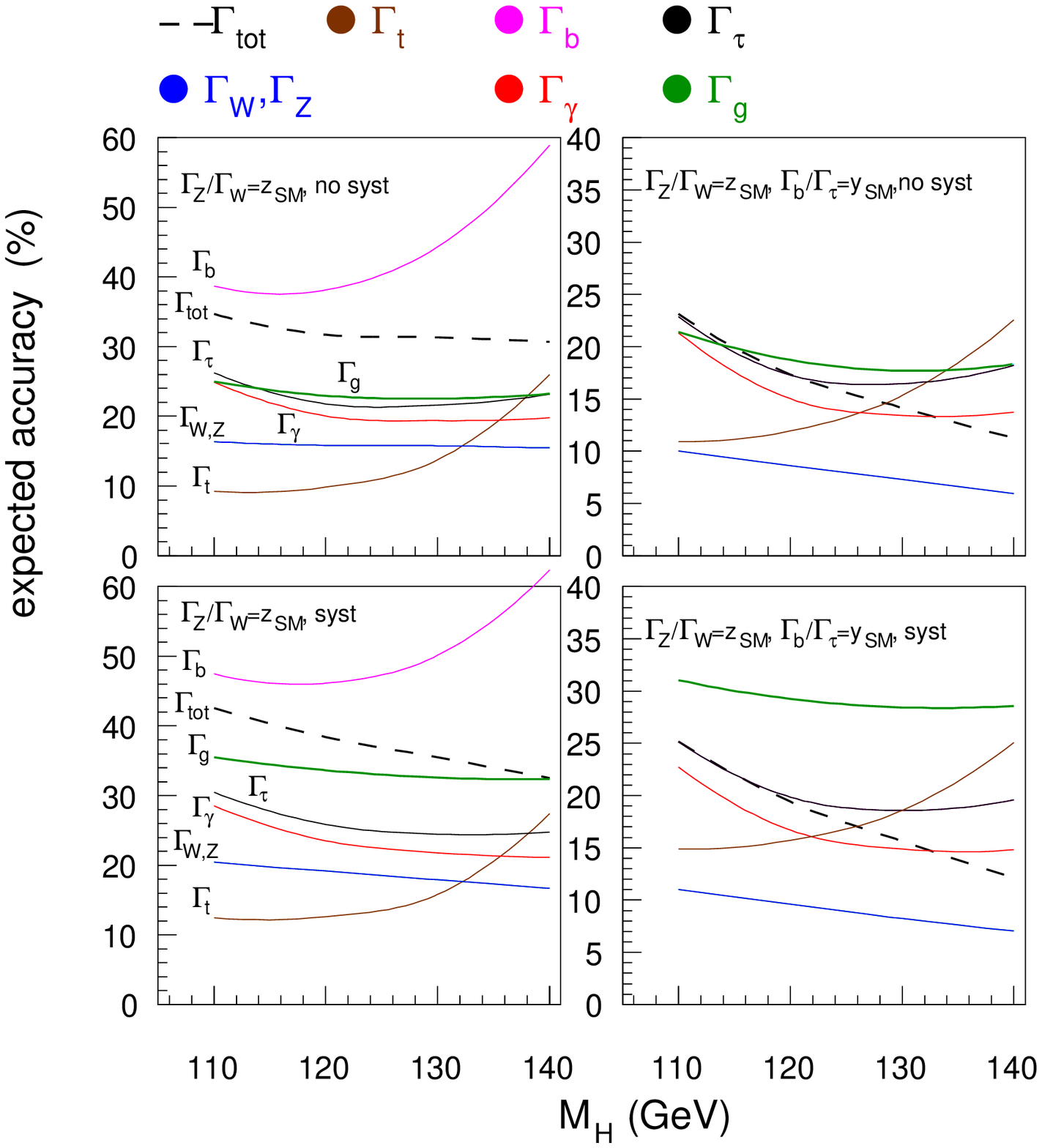}
}
\caption{\label{fig:fig1}
\small Relative accuracy (\%) on the individual rates 
$\Gamma_i$ expected at the LHC (from ref.~\cite{rb02}). 
See the text for a detailed description of the panels.}
\efig

Up to now detailed studies on signal and backgrounds for several channels 
have been performed, namely 
$gg \to H, (H \to \gamma \gamma, ZZ, 
WW)$~\cite{atl99-15,cms94-38,dreiner,denegri}, 
$qq \to qqH, (H \to $ $\gamma \gamma, \\ \tau \tau, 
WW)$~\cite{rz97,rzh99,prz00,rz99,kprz01}, 
$p p \to t \bar t H, (H \to b \bar b, WW, 
\tau \tau)$~\cite{rws99,dmd01,mrw02,rb02} and 
$p p \to W H, H \to b \bar b$~\cite{dmd02}.
 Each process depends on two\footnote{This is not true in the case
of the weak boson fusion channel where one needs to assume that the
ratio of $HWW$ and $HZZ$ is the same as in the SM} 
Higgs couplings, one from the Higgs boson production and one from the Higgs 
boson decay.

 Therefore, every production and
decay channel provides a measurement of the ratio $Z_j^{(i)} = 
\Gamma_i \Gamma_j / \Gamma$, where $i = g,W,t$ indicates the
particles involved in the production process while the index 
$j=b,\tau,W,Z,g,\gamma$  refers to the decay process. In case of 
$m_H < 140$~GeV, the above mentioned channels allow to 
express the individual
rates $\Gamma_t$, $\Gamma_b$, $\Gamma_\tau$, $\Gamma_W$, $\Gamma_g$ and
$\Gamma_\gamma$ as functions of the observables $Z_j^{(i)}$ and of the total 
Higgs width $\Gamma$~\cite{rb02}. 
With the assumption that the total 
width is saturated by the known channels $\Gamma = \Gamma_b + \Gamma_\tau 
+ \Gamma_W + \Gamma_Z + \Gamma_g + \Gamma_\gamma$ (otherwise new processes 
would be observed independently of any precision study), an expression for 
$\Gamma$ can be obtained in terms of the measured quantities 
$Z_j^{(i)}$~\cite{rb02}. 
Figure~\ref{fig:fig1}~\cite{rb02} summarizes the relative accuracy on the 
individual rates $\Gamma_i$ expected in the model-independent scenario as 
well as in a scenario with $\Gamma_b/\Gamma_\tau$ fixed to its SM value, 
assuming a total integrated luminosity of 200~fb$^{-1}$. 
The upper plots show the accuracies obtained without including any 
theoretical  error, while the lower plots show the same accuracies 
including a systematic theoretical error of $20\%$ for the $gg\to H$ channel, of
$5\%$ for the $qq\to qqH$, and of $10\%$ for the $pp\to t\bar{t}H$.
 As can be seen, the total Higgs width can be indirectly 
determined in the low mass region with a precision of the order of 30\% in a 
model-independent way while 
the Higgs couplings can be determined with 
accuracies between 7\% and 25\%. In the case of $140 < m_H < 200$~GeV, the 
gluon fusion, weak boson fusion and $t \bar t H$ associated production 
processes, with the Higgs 
boson decaying only to gauge bosons, allow an indirect determination of 
$\Gamma_W$ and $\Gamma$ with a precision of the order of 10\%~\cite{zknrw00,z02}. 
In this Higgs mass range, however, there is no handle to study the Higgs Yukawa 
coulings to $b$ quarks and $\tau$ leptons. 
The assumption $\Gamma_Z / \Gamma_W = z_{SM}$ 
can be tested at the 20--30\% level, for $m_H > 130$~GeV, 
by measuring the ratio $Z_Z^{(g)} / Z_W^{(g)}$~\cite{z02}, 
and it can even be tested with the same level of accuracy for lower 
Higgs boson masses by comparing the two ratios $Z_b^{(WH)} / Z_b^{(t)}$ and 
$Z_\tau^{(W)} / Z_\tau^{(t)}$~\cite{rb02}. For $m_H > 140$~GeV, 
with luminosities of the order of 
300~fb$^{-1}$, the ratio $\Gamma_t/\Gamma_g$ can be tested in a 
model-independent measuring 
$Z_W^{(t)}/Z_W^{(g)}$~\cite{mrw02}. 

\section{$\mathbf{H \to b \bar b}$ via Weak Boson Fusion}

To improve the analysis of the $H b \bar b$ Yukawa coupling, one can
consider the decay of an Higgs, produced via Weak Boson 
Fusion, into $b \bar b$ pairs~\cite{hbb}. 

Signal and background event estimates are based on a leading order 
partonic calculation of the matrix elements (ME) obtained with the 
event generator ALPGEN~\cite{alpgen}. 
The  background sources considered include:
\begin{enumerate}
\item   QCD production of $b\bar{b}jj$ final states, where $j$ indicates a jet originating from a
light quark ($u,d,s,c$) or a gluon;
\item QCD production of  $jjjj$ final states; 
\item associated production of $Z^*/\gamma^* \to b\bar{b}$ and light
jets
\end{enumerate}
along with multiple interaction events ($pp\oplus pp,pp\oplus pp\oplus pp...$) 
giving rise to final states of the kind 
$b\bar{b}jj$ and $jjjj$.
In order to satisfy the requirements of optimization of the signal 
significance ($S/\sqrt{B}$) and 
compatibility with trigger and data acquisition constraints,
different selection criteria have been considered.
The 
sensitivity can be as large as $5$ for Higgs 
masses close to the exclusion limit given by LEP searches
but the ratio $S/B$ is only a fraction of a
percent. This implies that the background will have to be known
with accuracies at the permille level.  
The
background should therefore be determined entirely from data. 
The large rate of $b \bar b j j$ from single and multiple interactions 
and the smoothness of their mass distribution 
 will allow to estimate their size with enough statistical accuracy, 
without significant systematic uncertainties. 

The situation is  different in
the case of the  backgrounds from the tails of the $Z$
decays. The $Z$ mass peak is sufficiently close to $m_H$
to possibly
distort the $m_{bb}$ spectrum and spoil the ability to accurately
reconstruct  the noise level from data.
These backgrounds rates are at most comparable to the signal at low $m_H$. 
A 10\% determination of these final states, which should be easily achievable
using the $(Z\to \ell^+\ell^-)jj$ control sample and folding in the
detector energy resolution for jets, should therefore be sufficient
to fix these background  with the required accuracy.

Concerning the multiple interactions, in the simplest case of two 
overlapping events ($pp\oplus pp$), 
there are four possible  (including mistagging effects) 
combinations of events leading 
to a $b\bar{b}jj$ background.
The largest 
contribution arises from  $(jj_b) \oplus (jj_b)$ events,
where the $b\bar{b}$ mass spectrum has a broad peak in the middle of
the signal region. The absolute rate of these events 
 can be determined if the distribution of the beam-line
$z$ vertex separation
between the two overlapping events can be determined with a resolution of
of 5-10~mm. The nunber of these events is significantly lowered
using the higher threshold of 80~GeV for the forward jets.

Table~\ref{tab:ssig60} summarizes the accuracy reachable in the 
${\cal B}(H \to b \bar b)$ and in the $H b \bar b$ Yukawa 
coupling for the case of two different event selections (described 
in detail in ref.~\cite{hbb}), 
assuming that the coupling $HWW$ is the one predicted by the Standard
Model or determined in other reactions studied in the
literature. 
An integrated luminosity of $600$~fb$^{-1}$ 
is considered. 
\begin{table}[h]
\begin{center}
\begin{tabular}{lllll}\hline
& $m_H$~(GeV) & 115  & 120  & 140 \\
\hline
$(a)$ & $\delta \Gamma_b/\Gamma$ & $0.33$  & $0.35$   & $0.71$ \\
 & $\delta y_{Hbb}/y_{Hbb}$ & $0.58$  & $0.51$   & $0.56$ \\
\hline
$(b)$ & $\delta \Gamma_b/\Gamma$ & $0.20$  & $0.19$   & $0.37$ \\ 
 & $\delta y_{Hbb}/y_{Hbb}$ & $0.36$  & $0.30$   & $0.29$ \\ 
\hline
\end{tabular}            
\caption{\label{tab:ssig60}
{\small The statistical significance of the determination of 
the branching ratio $\Gamma_b / \Gamma$ and of the
 $b$-quark Yukawa coupling 
in the configurations (a) and (b) (see ref.~\cite{hbb} for a  
description of the event selections), for an 
integrated luminosity of $600$~fb$^{-1}$. 
The $p_{\rm T}^j$ cut on jets 
is $p_{\rm T}^j > 60$~GeV. The case of $p_{\rm T}^j > 80$~GeV \cite{hbb}
 doesn't affect sizeably the results. 
The probability to mistag a light-jet as b-jet is assumed to be
$0.01$.}}
\end{center}
\end{table}
The $H\to b\bar{b}$ decay in the WBF channel could be used 
together with other processes already examined in the literature for a model
independent determination of the ratio of Yukawa couplings
$y_{Hbb}/y_{H\tau\tau}$~\cite{zepp}.

As a conclusion of the analysis presented in ref.~\cite{hbb}, 
the $H\to b\bar{b}$ channel produced in association 
with two jets is suggested as an additional channel to be 
exploited for the measurements of the Higgs couplings to fermions.

\section{Summary}
During the last few years there has been a dramatic improvement in both 
theoretical and experimental studies of several Higgs boson production 
and decay 
channels at the LHC. A strategy has been designed to study, in a
model-independent way, the Higgs couplings to fermions and bosons, which allows 
also, with little theoretical assumption, an indirect determination of the 
total Higgs width. The main results of a very recent analysis of the 
$H\to b \bar b$ channel in Weak Boson Fusion production have been reviewed, 
pointing out its importance for the determination of the $H b \bar b$ Yukawa 
coupling. 
\vskip3mm
\noindent
{\bf Acknowledgements}\\
I wish to acknowledge 
M.L.~Mangano, F.~Piccinini, R.~Pittau and A.~Polosa 
for fruitful collaboration. I wish to thank the organizers for 
the kind invitation and for the pleasant atmosphere during the Workshop.

%

\begin{thebibliography}{9}
\bibitem{atl99-15} ATLAS Collaboration, Tech. Rep. CERN/LHCC/99-15, CERN, 1999.
\bibitem{zknrw00} R.~Kinnunen {\it et al.}, Phys.~Rev.~D62 (2000) 013009. 
\bibitem{grazzini} M.~Grazzini, hep-ph/0209302; 
S.~Catani, D.~de~Florian and M.~Grazzini, JHEP 0105 (2001) 025; and 
JHEP 0201 (2002) 015;
G.~Bozzi, S.~Catani, D.~de Florain, M.~Grazzini, hep-ph/0302104; 
S.~Catani, D.~de Florain, M.~Grazzini, P.Nason, hep-ph/0306211.
\bibitem{kilgore} W.B.~Kilgore, these proceedings; 
R.V.~Harlander and W.B.~Kilgore, Phys.~Rev.~D64 (2001) 025; 
Phys.~Rev.~Lett.~88 
(2002) 201801; C.~Anastasiou and K.~Melnikov, \npb~646 (2002) 220; 
V.~Ravindran, J.~Smith, W.L.~Van Neerven, hep-ph/0302135.
.
\bibitem{tthnlo} W.~Beenakker {\it et al.}, Phys. Rev. Lett. 87 (2001) 201805; 
                 S.~Dawson and L.~Reina, Phys. Rev. Lett. 87 (2001) 201804; 
            S.~Dawson, L.~Reina and D.~Wackeroth, Phys. Rev. D65 (2002) 053017;
S.~Dawson, C.~Jackson, L.H.~Orr, L.~Reina and D.~Wackeroth, hep-ph/0305087.
\bibitem{diphox} T.~Binoth {\it et al.}, Eur. Phys. J. C16 (2000) 311; 
hep-ph/0203064.
\bibitem{gluglugg} Z.~Bern, L.~Dixon and C.~Schmidt, hep-ph/0206194.
\bibitem{evc} R.K.~Ellis and S.~Veseli, Phys.~Rev.~D60 (1999) 011501; 
J.~Campbell, hep-ph/0105226.
\bibitem{ce} J.~Campbell and R.K.~Ellis, hep-ph/0202176.
\bibitem{fw} S.~Frixione and B.R.~Webber, hep-ph/0204244. 
\bibitem{qcdhjj} V.~Del~Duca {\it et al.}, Nucl.~Phys.~B616 (2001) 367.
\bibitem{alpgen} M.L.~Mangano {\it et al.}, hep-ph/0206293.
\bibitem{comphep} A.~Pukhov {\it et al.}, hep-ph/9908288.
\bibitem{grace} T.~Ishikawa {\it et al.}, MINAMI-TATEYA Group Coll., KEK-92-19;
\bibitem{madgraph} T.~Stelzer and W.F.~Long, Comput.~Phys.~Commun.~81 (1994) 357; 
F.~Maltoni and T.~Stelzer, hep-ph/0208156.
\bibitem{acermc} B.P.~Kersevan and E.~Richter-Was, hep-ph/0201302.
\bibitem{cms94-38} CMS Collaboration, Tech. Rep. CERN/LHCC/94-38, CERN, 1994.
\bibitem{dreiner} M. Dittmar, and H.K. Dreiner, Phys.~Rev.~D55 (1997) 167; 
hep-ph/9703401.
\bibitem{denegri} D.~Denegri {\it et al.}, hep-ph/0112045.
\bibitem{rz97} D.~Rainwater and D.~Zeppenfeld,  JHEP 12 (1997) 005.
\bibitem{rzh99} D.~Rainwater, D.~Zeppenfeld and K.~Hagiwara, Phys. Rev. D59
(1999) 014037.
\bibitem{prz00} T.~Plehn, D.~Rainwater and D.~Zeppenfeld, 
Phys. Rev. D61 (2000) 
\bibitem{rz99} D.~Rainwater and D.~Zeppenfeld, Phys. Rev. D60 (1999) 113004.
\bibitem{kprz01} N.~Kauer {\it et al.}, Phys. Lett. B503 (2001) 113.
\bibitem{rws99} E.~Richter-Was and M.~Sapinski, Acta Phys. Polon. B30 (1999)
1001. 
\bibitem{dmd01} V.~Drollinger, T.~M\"uller and D.~Denegri, hep-ph/0111312.
\bibitem{mrw02} F.~Maltoni, D.~Rainwater and S.~Willenbrock, hep-ph/0202205.
\bibitem{rb02} A.~Belyaev and L.~Reina, hep-ph/0205270.
\bibitem{dmd02} V.~Drollinger, T.~M\"uller and D.~Denegri, hep-ph/0201249.
\bibitem{z02} D.~Zeppenfeld, hep-ph/0203123.
\bibitem{hbb} M.L.~Mangano {\it et al.}, hep-ph/0210261.
\bibitem{zepp} D. Rainwater, D. Zeppenfeld and K. Hagiwara, Phys. Rev.
{\bf D59}, 014037 (1999).
\end{thebibliography}
%
%
%

\end{document}